\documentclass[aip,apl,amsmath,amssymb,reprint,floatfix]{revtex4-1}

\usepackage{hyperref}
\usepackage{graphicx}
\usepackage{times}

\newcommand{\rnd}[1]{\left( #1  \right)}
\newcommand{\of}[1]{\!\left( #1  \right)}
\newcommand{\sqr}[1]{\left[ #1  \right]}
\newcommand{\unit}[1]{\;\text{#1}}
\newcommand{\micron}{\mu\text{m}}

\begin{document}

\title{Blowing of Polysilicon Fuses} \thanks{Copyright (2010) American
  Institute of Physics. This article may be downloaded for personal
  use only. Any other use requires prior permission of the author and
  the American Institute of Physics.The following article appeared in
  \textit{Appl.\ Phys.\ Lett.\ }\textbf{97}:023502 (2010) and may be
  found at \url{http://link.aip.org/link/?apl/97/023502}}

\newcommand{\ulAddress}{MACSI, Department of Mathematics and
  Statistics,  University of Limerick, Limerick, Ireland}
\newcommand{\analogAddress}{Analog Devices, Raheen Industrial
    Estate,  Limerick, Ireland.}

\author{W. T. Lee }
\author{A. C. Fowler}
\author{O. Power}
\affiliation{\ulAddress}
\author{S. Healy}
\author{J. Browne}
\affiliation{\analogAddress}



\begin{abstract}
Polysilicon fuses are one time programmable memory elements which
allow the calibration of integrated circuits at wafer and package
level. We present a zero dimensional lumped parameter model of the
programming of fuses made from a combination of tungsten silicide and
polycrystalline silicon. The components of the model are an electrical
model, a thermal model and a flow model. The electrical model
describes the temperature and geometry dependent resistance of the
fuse. The thermal model describes the heating and melting of the fuse
and its surroundings. The flow model describes the disconnection of
the fuse by electromigration driven flow of silica.  The model
generates quantitatively accurate results and reproduces trends with
applied voltage and fuse size.
\end{abstract}

\preprint{MACSI/WTL/1}

\maketitle


Polysilicon fuses are one time programmable memory elements in silicon
chips. They are utilised as permanent on chip memory storage for
calibration factors in precision analogue circuits and can be
programmed (blown) at both wafer and package levels. Polysilicon fuses
are constructed from layers of polycrystalline silicon (polysilicon)
and metal silicides. Theories of the blowing process which have been
put forward invoke grain recrystallisation~\cite{Tonti2003}, melting
and vapourisation~\cite{Li2006} and electromigration~\cite{Doorn2007}.

\begin{figure}
\begin{center}
\includegraphics{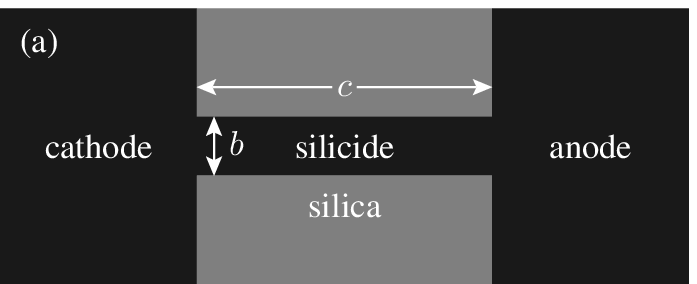}\\
\includegraphics{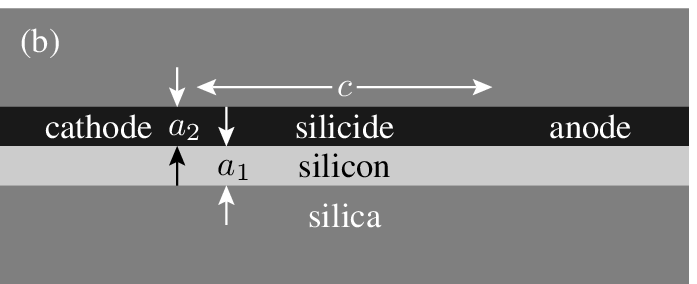}
\end{center}
\caption{\label{fuse_geometry} Polysilicon fuse geometry. (a) Top
  view. (b) Transverse cross-section.}
\end{figure}

One fuse design is shown in Figure~\ref{fuse_geometry}. The fuse is
completely solid state and consists of layers of conducting tungsten
silicide and semiconducting polysilicon surrounded by insulating
silica. The thickness of the polysilicon and tungsten silicide layers
are $a_1$ and $a_2$ respectively, the width of the fuse is $b$ and its
length is $c$. The electrodes on either side of the fuse are made from
the same materials with the same thickness but are much larger in the
other two directions.

\begin{figure}
\begin{center}
\includegraphics{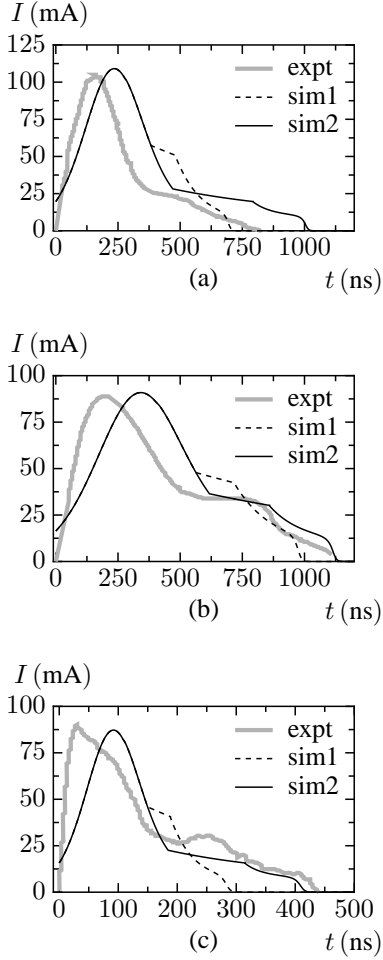}
\end{center}
\caption{\label{current_time_data} Current time curves for: (a)
  $6\unit{V}$ applied to a $0.3\;\micron \times 0.6\;\micron \times
  3\;\micron$ fuse; (b) $5\unit{V}$ applied to a $0.3\;\micron \times
  0.6\;\micron \times 3\;\micron$ fuse; (c) $4.8\unit{V}$ applied to a
  $0.3\;\micron \times 0.3\;\micron \times 1.5\;\micron$ fuse. Grey
  lines, experimental data. Solid black line, simulations including
  superheating. Dashed black line, simulations without
  superheating. Experimental curves in (a) and (b) were calculated
  from data in Ref.~\onlinecite{Li2006}.}
\end{figure}

\begin{figure}
\begin{center}
\includegraphics{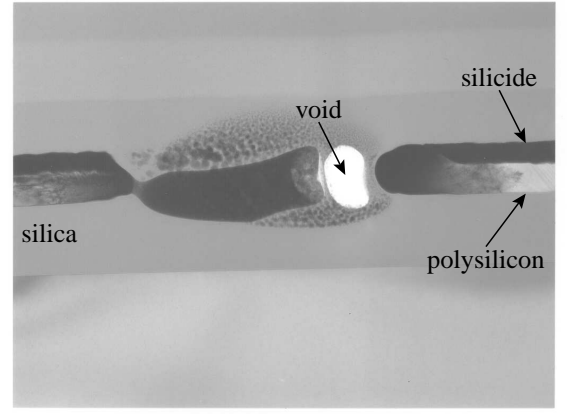}
\end{center}
\caption{\label{blown_microstructure} Electron micrograph showing the
  microstructure of a blown $0.3\;\mu\text{m} \times 0.3\;\mu\text{m}
  \times 1.5\; \mu\text{m}$ fuse.}
\end{figure}

The fuse is blown by the application of a fixed potential bias across
the fuse. Current versus time graphs are shown for different voltages
and different size fuses in Figure~\ref{current_time_data}. The
microstructure of a blown $0.3\;\micron \times 0.3\;\micron \times
1.5\;\micron$ fuse is shown in Figure~\ref{blown_microstructure}. The
following features of the blowing process are apparent: (i) there is
evidence of melting of the fuse, and of parts of the electrodes and
the silica surrounding the fuse; (ii) the fuse is disconnected by both
a void and by a layer of silica. A model of the fuse blowing process
must explain these observations.

We propose that the fuse blowing process occurs by the following
steps.  (1)~When the current starts to flow, it flows mostly through
the metallic silicide. As the fuse heats up, the silicide conductivity
reduces but the conductivity of the semiconducting polysilicon
increases until it is the dominant conductor. The overall effect is
that the conductivity of the fuse first increases then
decreases. (2)~Eventually the fuse, parts of the electrodes and some
of the silica surrounding the fuse melt.  (3)~Electromigration drives
the fuse material towards the cathode. This displaces the insulating
molten silica towards the anode, where it begins to pinch off the fuse
from the anode.  (4)~When pinch off is nearly complete the fuse
ruptures forming the void (this process is driven by the contraction
of silicon on melting).

\begin{figure}
\begin{center}
\includegraphics{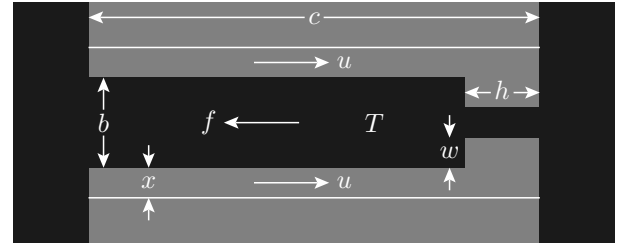}
\end{center}
\caption{\label{fuse_variables} Fuse variables: the temperature $T$ of
  the fuse and the width, $w$, and height $h$ of the pinched off
  region.}
\end{figure}

To model the fuse blowing process we use a lumped parameter approach.
This model builds on work reported in Ref.~\onlinecite{esgi62}. The
lumped parameter model of the fuse consists of ordinary differential
equations for the temperature, $T$, of the fuse, and the height, $h$,
and width, $w$, of the pinched off region. These variables are
illustrated in Figure~\ref{fuse_variables}. An electronic model describes
the temperature and geometry dependent resistance of the fuse; a
thermodynamic model describes the heating and melting of the fuse and
its surroundings; a flow model describes the electromigration driven
pinch off of the fuse. The rupture of the fuse is not modelled
explicitly: it is assumed that rupture occurs just as pinch off is
complete.

\paragraph{Electronics}
\newcommand{\Rt}{R_{\text{t}}}
\newcommand{\Rf}{R_{\text{f}}}
Electrically the system consists of the fuse in series with a
transistor. The fuse blowing process is initiated by a decrease of the
transistor resistance from a very large value to $\Rt$. This decrease
occurs on a nanosecond timescale and is assumed to be instantaneous in
the model. The fuse can be modelled as two resistors in series, with a
total resistance $\Rf$.  The first resistor represents the part of the
fuse unaffected by flow and has a temperature dependent
resistance, $R_1$. The second resistor represents the region of the
fuse in which pinch off occurs and thus has a temperature and geometry
dependent resistance, $R_2$.
\begin{align}
R_1&=\dfrac{c}{\sigma\of{T}ab}\\
R_2&=\dfrac{h}{\sigma\of{T}a\rnd{b-2w}}
\end{align}
The current $I$ flowing through the fuse is calculated from Ohm's law. 

Following Ref.~\onlinecite{Minixhofer2003}, conductivity, $\sigma\of{T}$, is
taken to be a quadratic function of temperature. There is insufficient
data to separate out the contributions of the polysilicon and tungsten
silicide layers, therefore a single conductivity function is used for
both layers of the fuse.  \newcommand{\Tmax}{T_{\text{max}}}
\begin{equation}
\sigma\of{T}=\sigma_0\sqr{1-\alpha\rnd{T-\Tmax}^2}
\end{equation}
As the fuse heats up, its conductivity increases until $\Tmax$ is
reached and then it decreases.

\paragraph{Thermodynamics}
The temperature, $T$, of the fuse is assumed to be homogeneous throughout
the fuse. The equation modelling its evolution is 
\begin{equation}
v C\of{T} \dfrac{\text{d}T}{\text{d}t}=I^2 \Rf  - \dfrac{k_3 s \rnd{T-T_0}}{d}
\end{equation}
The left hand side describes the heat capacity of the fuse. The terms
on the right hand side describe Joule heating and Newtonian cooling
respectively.  $C\of{T}$ is a the volume averaged heat capacity per
unit volume of the fuse calculated from: $C_{1}$ and $C_2$, the heat
capacity per unit volume of polysilicon and tungsten silicide (assumed
to be the same in the solid and liquid); and $\Delta H_1$ and $\Delta
H_2$, the enthalpies of fusion per unit volume of polysilicon and
tungsten silicide. In solving the equations, the melting phase
transitions (at $T_{\text{m}1}$ and $T_{\text{m}2}$ in polysilicon and
tungsten silicide respectively) are assumed to have a finite width of
$\delta T$. $v$ is the volume of the fuse, $k_3$ is the thermal
conductivity of silica, $T_0$ is room temperature and $s$ is the
surface area of the fuse, $2c\rnd{a_1+a_2+b}$, $d$ is a thermal
conduction length. A better fit to the data is obtained by allowing
superheating of the tungsten silicide before it melts: this seems
physically reasonable given the small size of the fuse and the rapid
heating rates. Superheating is included within the model by increasing
$T_{\text{m}2}$ from its equilibrium value of $2300\unit{K}$.

\paragraph{Fluid mechanics}

The body force driving electromigration is given by $f=g n_q I \Rf /
c$ where $n_q$ is the charge density of charge carriers and $g$ is a
constant of proportionality. The force acts on the conducting silicon
and silicide and is directed from the anode to the cathode. The
viscosities of the molten silicon and silicide are much lower than
that of molten silica. Therefore flow is limited by the motion of
silica. The electromigration body force sets up a pressure gradient
within the fuse which is transmitted to the molten silica surrounding
the fuse and causes it to move with a velocity $u=fx^2/12\eta$ where
$\eta$ is the dynamic viscosity of molten silica and $x$ is the width
of the molten silica region surrounding the fuse.

The equations describing the pinch-off of the fuse are
\begin{align}
\dfrac{\text{d}h}{\text{d}t}&=\dfrac{ux}{w}-\dfrac{fh^4}{\eta w^2}\\
\dfrac{\text{d}w}{\text{d}t}&=\dfrac{fh^3}{\eta w}
\end{align}
The first equation expresses conservation of volume, the second is a
crude balance between work done by body forces and work done against
friction in the silica. Flow is only allowed to occur once
$T_{\text{m}2}$ has been exceeded. To allow the equations to be solved
numerically $h$ and $w$ are initialised with small non-zero values.

The equations were solved numerically using a 4th order Runge-Kutta
scheme with adaptive time-stepping. Parameter values used in the
simulations are summarised in Table~\ref{parameters}. The results are
shown in Figure~\ref{current_time_data}, for models using the equilibrium melting value of $T_{\text{m}2}$ and the values including superheating given in Table~\ref{parameters}. Considering the simplicity of the model the results are
remarkably good and trends with voltage and fuse size are correctly
predicted.


A zero dimensional, lumped parameter model of the programming of
polysilicon fuses successfully reproduces the available experimental
data for one type of fuse. The model captures trends with voltage and
size. This strongly suggests that the assumptions used to construct
the model are fundamentally correct. To take the model further, more
experimental data are required, describing the temperature dependent
conductivity of the individual polysilicon and tungsten silicide
layers for example. This will allow three dimensional modelling of the
partial differential equations describing the system.

\begin{table}[!h]
\caption{\label{parameters}Values of parameters used in simulations.}
\begin{center}
\begin{tabular}{c r@{$\;$}l c}
\hline
Parameter  &  \multicolumn{2}{c}{Value}   \\
\hline \hline
$\Rt$              &   5&$\Omega$  \\
$T_{\text{m}1}$    &   1700&$\text{K}$  \\
$\eta$             &   10&$\text{Pa}\unit{s}$  \\
$n_q$               &   $3.16\times 10^{10}$&$\text{C}\unit{m}^{-3}$ \\
$g$                &   0.2&\\
$\sigma_0$         &   3.33&$\Omega^{-1}\unit{m}^{-1}$\\
$\Tmax$            &   1400&$\text{K}$\\
$T_0$            &   300&$\text{K}$\\
$\alpha$           &   $6.89\times 10^{-7}$&$\text{K}^{-2}$\\
$C_1$ & $1.66\times 10^6$&$\text{J}\unit{m}^{-3}\unit{K}^{-1}$ \\
$C_2$              &   $3.09\times 10^8$&$\text{J}\unit{m}^{-3}\unit{K}^{-1}$\\
$\Delta H_1$       &   $4.15\times 10^9$&$\text{J}\unit{m}^{-3}$\\
$\Delta H_2$       &   $8.30\times 10^{10}$&$\text{J}\unit{m}^{-3}$\\
$\delta T$         &   $50$&$\text{K}$ \\
$k_3$              &   $1.5$&$\text{W}\unit{m}^{-1}\unit{K}^{-1}$ \\
\hline
\end{tabular}

\vspace{4mm}

\begin{tabular}{c r@{$\;$}l r@{\;}l r@{\;}l}
\hline
Property   &  \multicolumn{2}{c}{(a)}  &  \multicolumn{2}{c}{(b)}  
                                          &  \multicolumn{2}{c}{(c)}  \\
\hline  \hline
$a_1$ & $0.15$&$\mu\text{m}$ & $0.15$&$\mu\text{m}$ & $0.15$&$\mu\text{m}$ \\
$a_2$ & $0.15$&$\mu\text{m}$ & $0.15$&$\mu\text{m}$ & $0.15$&$\mu\text{m}$ \\
$b$   & $0.6$&$\mu\text{m}$ & $0.6$&$\mu\text{m}$ & $0.3$&$\mu\text{m}$ \\
$c$   & $3$&$\mu\text{m}$ & $3$&$\mu\text{m}$ & $1.5$&$\mu\text{m}$ \\
$d$ &   $0.3$&$\mu\text{m}$ & $0.3$&$\mu\text{m}$ & $0.15$&$\mu\text{m}$ \\
$x$ &   $0.1$&$\mu\text{m}$ & $0.1$&$\mu\text{m}$ & $0.08$&$\mu\text{m}$ \\  
$V$ &   $5$&$\text{V}$ & $6$&$\text{V}$ & $4.8$&$\text{V}$ \\
$T_{\text{m}2}$ & $2400$&$\text{K}$ & $2500$&$\text{K}$ & $2500$&$\text{K}$ \\
\hline
\end{tabular}
\end{center}
\end{table}

\begin{acknowledgments}
We acknowledge support of the Mathematics Applications Consortium for
Science and Industry (www.macsi.ul.ie) funded by the Science
Foundation Ireland Mathematics Initiative Grant 06/MI/005.
\end{acknowledgments}


%

\end{document}